Suppression of superconductivity by Sb-substitution into $CeOBiS_2$ single crystals

Tatsuya Suzuki[1], Yuji Hanada[1], Masanori Nagao[1*], Yuki Maruyama[1], Satoshi Watauchi[1], and Isao Tanaka[1]

[1]*University of Yamanashi, 7-32 Miyamae, Kofu, Yamanashi 400-0021, Japan*

*Corresponding Author

Masanori Nagao

Postal address: University of Yamanashi, Center for Crystal Science and Technology

Miyamae 7-32, Kofu, Yamanashi 400-0021, Japan

Telephone number: (+81)55-220-8610

Fax number: (+81)55-220-8270

E-mail address: mnagao@yamanashi.ac.jp

**Highlights**

We have successfully grown Sb-doped CeOBiS$_2$ single crystals.

Superconductivity in CeOBiS$_2$ disappeared by 4-5 atomic percent Sb-doping.

The correlations between Ce valence state and superconductivity were not observed.

**Abstract**

Sb-substituted CeOBiS$_2$ single crystals with 0.2-1.0 mm size have been successfully grown using CsCl/KCl flux. Sb substitution strongly suppressed the superconductivity in CeOBiS$_2$, and the substitution of more than approximately 4 at% in Bi-site disappeared the superconductivity at a measurement range of above 0.36 K. Furthermore, the drastic change of *c*-axis lattice parameter was observed in a neighborhood to Sb substitution amount, which maybe suggests the structural transition in CeOBiS$_2$. The Ce valence state of the obtained single crystals was evaluated. However, the established correlations between Ce valence state and superconductivity could not be revealed.

**Main text**

**Introduction**

Layered compounds often exhibited superconductivity with high transition temperatures, which were commencing with cuprate superconductors [1,2] and iron-based superconductors [3]. $BiS_2$-based layered superconductors were formed from superconducting layer and block layers which were similar layered structures to iron-based superconductors [4]. $La(O,F)BiS_2$ and $La(O,F)FeAs$ were one of the $BiS_2$-based and iron-based superconductors, respectively. The difference of those superconductors was only superconducting layers which are between $BiS_2$-layer and FeAs-layer. However, the superconducting transition temperature of $La(O,F)BiS_2$ superconductors was extremely lower than that of $La(O,F)FeAs$ superconductors. The clarification of that reason is important for the mechanism of high transition temperature superconducting phenomena. We focused on $La(O,F)BiS_2$ which is a typical $BiS_2$-based superconductor. La-site in $La(O,F)BiS_2$ can be substituted by each/several rare earth elements (*RE*: Ce, Pr, Nd, Sm, Yb) [5-11]. The O-site was partially substituted by F in the block layers of those compounds which were *RE*(O,F)-layers. In consequence, the electron carriers were supplied to the $BiS_2$-layer (superconducting layer), resulting in the inducement of superconductivity. Ag [12], Cu [13], Pb [14], and Cd [15] were performed to substitute with partially Bi-site, or S-site was substituted by Se [16,17] in the $BiS_2$-layer which was the superconducting layer. Especially, Pb and Se substitutions enhanced the superconducting transition temperature. In contrast, La-site perfectly/partially substituted with Ce ($CeOBiS_2$) exhibited superconductivity without F-substitution [18-21]. The origin of this superconductivity was the supply of carriers from the valence fluctuation of Ce [22]. On the other hand, The Bi-site can be perfectly replaced by Sb, which was a $CeOSbS_2$ compound. Despite this compound was observed Ce valence fluctuations, it exhibited no superconductivity [23]. Moreover, the crystal system becomes the change between $CeOBiS_2$ and $CeOSbS_2$. $CeOBiS_2$ and $CeOSbS_2$ were tetragonal and monoclinic crystal systems, respectively. In focusing on other superconductors, Sb substitution in Bi-site suppressed the superconductivity such as $RhX_2$ [24] and $LaNiX_2$ ($X$ = Bi, Sb) [25]. Sb substituted with Bi-site in superconductors may be an important role in superconductivity.

In this paper, a systematic investigation between Sb substitution and superconductivity for $CeOBiS_2$ superconductor was performed using single crystals grown by CsCl/KCl flux method [26,27]. The relationship between Sb substitution and Ce valence state was also investigated. The boundary Sb concentration of superconductivity was revealed in $CeO(Bi,Sb)S_2$ single crystals which were partially substituted with Sb in $CeOBiS_2$.

**Methods**

Sb-substituted CeOBiS$_2$ single crystals were grown using CsCl/KCl flux [18,23]. The raw materials of Ce$_2$S$_3$, Bi$_2$O$_3$, Bi$_2$S$_3$, Sb$_2$O$_3$, Sb$_2$S$_3$ were weighed to have a nominal composition of CeOBi$_{1-x}$Sb$_x$S$_2$ ($0.02 \leq x \leq 0.40$). The molar ratio of the CsCl/KCl flux was CsCl:KCl = 5:3. The mixture of the raw materials (0.8 g) and CsCl/KCl flux (5.0 g) was ground using a mortar and then sealed in an evacuated quartz tube (~10 Pa). The prepared quartz tube was heated at 950 °C for 10 h, followed by cooling to 600 °C at a rate of 1 °C/h, then the sample was cooled down to room temperature (~30 °C) in the furnace. The heated quartz tube was opened to air and the obtained materials were washed and filtered to remove the CsCl/KCl flux using distilled water.

The compositional ratio of the grown crystals was evaluated using energy dispersive X-ray spectrometry (EDS) (Bruker; Quantax 70) associated with the observation of the microstructure, based on scanning electron microscopy (SEM) (Hitachi High-Technologies; TM3030). Analytical compositions of each element were defined as $C_{XX}$ (*XX*: Ce, O, Bi, Sb, S). The obtained compositional values were normalized using $C_{Bi} + C_{Sb} = 1.0$ (Bi + Sb analytical composition was 1.0), and then Ce, O, and S compositions ($C_{Ce}$, $C_O$, and $C_S$) were determined. The crystal structure and the *c*-axis lattice parameters of the grown crystals were identified using X-ray diffraction (XRD) (Rigaku; MultiFlex) with CuK$\alpha$ radiation.

The resistivity-temperature ($\rho$-$T$) characteristics of the grown single crystals were determined using the standard four-probe method with a constant current density ($J$) mode using a physical property measurement system (Quantum Design; PPMS DynaCool). The electrical terminals were fabricated using Ag paste (DuPont; 4922N). The $\rho$–$T$ characteristics in the temperature range of 0.36-15 K were determined based using the adiabatic demagnetization refrigerator (ADR) option for the PPMS. A magnetic field of 3.0 T at 1.9 K was utilized to operate the ADR, which was subsequently removed. Consequently, the temperature of the sample decreased to approximately 0.36 K. The measurement of $\rho$–$T$ characteristics were initiated at the lowest temperature (~0.36 K), which was spontaneously increased to 15 K. The superconducting transition temperature ($T_c$) was estimated from the resistivity measurement. The zero resistivity ($T_c^{zero}$) is determined as the temperature at which the resistivity is below approximately 100 μΩ cm. The obtained resistivity values were normalized by $\rho/\rho_{15K}$. ($\rho_{15K}$: the resistivity value at 15 K). The transition temperature corresponding to the onset of superconductivity ($T_c^{onset}$) is defined as the temperature at which deviation from linear behavior is observed in the normal conducting state for the temperature dependence of normalized resistivity value ($\rho/\rho_{15K}$–$T$).

The valence state of the Ce and Bi component in the grown single crystals was estimated using X-ray absorption fine structure (XAFS) spectroscopy analysis of Ce-$L_3$ and Bi-$L_3$ edges using an Aichi XAS beamline with synchrotron X-ray radiation (BL5S1: Proposal No. 202104005 and 2022D3006). To prepare samples for the XAFS spectroscopy measurements, the obtained single crystals were ground and mixed with boron nitride (BN) powder, before being pressed into a pellet of 4 mm diameter. The ratios of $Ce^{3+}$ and $Ce^{4+}$ in the grown single crystals were calculated using linear combination fitting of the $Ce_2S_3$($Ce^{3+}$) and $CeO_2$($Ce^{4+}$) reference samples. For the Bi valence state, reference samples were $Bi_2S_3$($Bi^{3+}$) and $NaBiO_3$($Bi^{5+}$).

**Results and Discussion**

Figure 1 shows a typical SEM image of a CeO(Bi,Sb)$S_2$ single crystal which was a plate-like shape with a size of 0.2-1.0 mm and a thickness of approximately 150 μm. Table I shows the nominal Sb contents ($x$) and analytical compositions of each element (Ce, O, Bi, Sb, and S) for the obtained crystals. Analytical Sb compositions ($C_{Sb}$) were lower than the nominal Sb contents ($x$). The $C_{Sb}$ values of the obtained crystals with $x$ = 0.05-0.08 exhibited similar values within the analytical error using EDS analysis. The values of analytical compositions for other elements ($C_{Ce}$, $C_O$, and $C_S$) were higher than the stoichiometric compositions which were Ce:O:S = 1:1:2. In other words, the total of the Bi and Sb compositions was lower than the stoichiometric composition. It suggested that Bi-site becomes a slight deficiency. Previous reports indicated the Bi-site deficiency in the parent compounds [28,29].

Figure 2 (a) shows XRD patterns of a well-developed plane in the grown CeO(Bi,Sb)S$_2$ single crystals. The obtained XRD patterns corresponded to only (00$l$) diffraction peaks of CeOBiS$_2$ structure [18]. It indicated that CeO(Bi,Sb)S$_2$ single crystals were grown and the $c$-plane is well-developed. The $c$-axis lattice parameters were calculated from (00$\underline{11}$) diffraction peaks. Analytical Sb compositions ($C_{Sb}$) dependence of the $c$-axis lattice parameters was shown in Figure 2 (b). The $c$-axis lattice parameters were almost decreased with increasing $C_{Sb}$ except for $C_{Sb} = 0$. It indicated that the Bi-site was substituted by Sb which is small ionic radii. However, the $c$-axis lattice parameters of the non-Sb substitution sample ($C_{Sb} = 0$) were smaller than those of $C_{Sb} \leq 0.043$ samples. That reason is unclear. The decreasing rate of the $c$-axis lattice parameters was drastically changed at the boundary between $C_{Sb} = 0.043$ and 0.045. We presumed that the crystal system was changed to a monoclinic crystal system. CeOBiS$_2$ is tetragonal crystal system [22]. On the other hand, CeOSbS$_2$ exhibited the monoclinic crystal system [23], which is completely Sb substitution in Bi-site. Approximately 5 at% Sb substitution in the Bi-site may change the crystal system of $CeOBiS_2$.

The superconductivity in various Sb substituted $CeO(Bi,Sb)S_2$ single crystals was investigated. Figure 3 (a) shows the temperature dependence of the normalized at 15 K resistivity ($\rho/\rho_{15K}$) for the grown $CeO(Bi,Sb)S_2$ single crystals. Superconducting transitions were observed in the $CeO(Bi,Sb)S_2$ single crystals with $C_{Sb}$ = 0.015, 0.023, 0.035, and 0.043 at above 0.36 K. The resistivity in the normal state (above the $T_c$) was slightly decreased with an increasing temperature, exhibiting the semiconducting behavior. On the other hand, the single crystals with $C_{Sb} \geq 0.045$ showed non-superconductivity with only semiconducting behavior. Therefore, we simply presumed that the carrier concentration of $CeO(Bi,Sb)S_2$ has decreased with the increase in Sb substitution. In the previous report [22], $CeOBiS_2$ superconductivity originated from the carrier supply which was provided from the Ce mixed valence state ($Ce^{3+}$ and $Ce^{4+}$). The investigation of the Ce valence state in the grown $CeO(Bi,Sb)S_2$ single crystals will be shown in Figure 4. The $C_{Sb}$ dependence of the superconducting transition temperature ($T_c^{onset}$ and $T_c^{zero}$) was summarized in Figure 3 (b). The Sb substitution drastically decreased the superconducting transition temperature compared to the single crystal without Sb substitution ($C_{Sb}$ = 0). It indicates that Sb substitution strongly suppresses the superconductivity in CeOBiS$_2$. CeO(Bi,Sb)S$_2$ single crystals with $C_{Sb}$ ≤ 0.043 exhibited superconductivity. However, the $C_{Sb}$ values range of 0.043-0.045 was the same within the analytical error (See in Table I). Therefore, we considered that the boundary of superconductivity and non-superconductivity for the Sb substitution amount in CeOBiS$_2$ was approximately 0.04 (approximately 4 at% Sb substitution in the Bi-site).

Figure 4 (a) shows the Ce $L_3$-edge XAFS spectra at room temperature of some grown CeO(Bi,Sb)S$_2$ single crystals, Ce$_2$S$_3$ (Ce$^{3+}$ reference sample), and CeO$_2$ (Ce$^{4+}$ reference sample). The obtained Ce $L_3$-edge XAFS spectra demonstrated a peak at around 5725 eV and were assigned to trivalent electronic configuration (Ce$^{3+}$) [30]. Moreover, the peaks around 5730 eV and 5737 eV were assigned to tetravalent electronic configuration (Ce$^{4+}$) [31]. All evaluated CeO(Bi,Sb)S$_2$ single crystals exhibited a Ce valence fluctuation caused by a mixture state of Ce$^{3+}$ and Ce$^{4+}$. The ratios of Ce$^{3+}$ and Ce$^{4+}$ were estimated using linear combination fitting of Ce$_2$S$_3$ (Ce$^{3+}$) and CeO$_2$ (Ce$^{4+}$) through XAFS spectra. Ce$^{4+}$ ion concentration ratios of each analytical Sb composition ($C_{Sb}$) for the grown CeO(Bi,Sb)S$_2$ single crystals were shown in Figure 4 (b). The Ce$^{4+}$ ion concentration ratios were increased with the increase $C_{Sb}$, except for the $C_{Sb}$ = 0. However, no clear signature difference was observed at the boundary between superconductivity and non-superconductivity corresponding to $C_{Sb}$ = 0.043 and 0.045, respectively. Therefore, there was no clear correlation of $Ce^{4+}$ ion concentration ratios with $C_{Sb}$. It indicates that the superconductivity in CeO(Bi,Sb)S$_2$ may not be affected by the Ce mixed valence state. On the other hand, the Bi valence state of $Bi^{3+}$ and $Bi^{5+}$ for the grown CeO(Bi,Sb)S$_2$ single crystals was evaluated using Bi $L_3$-edge XAFS spectra. Only the $Bi^{3+}$ state was observed in all evaluated samples which were the same samples in Figure 4. The Sb valence state was not able to be evaluated by an Aichi XAS beamline for the X-ray energy. Further characterizations of these single crystals are therefore necessary to clarify the superconductivity suppression from the Sb substitution. Additional experiments such as Sb valence state and precise single crystal structure analysis are also required.

**Conclusions**

CeO(Bi,Sb)S$_2$ single crystals with a size of 0.2-1.0 mm and approximately 150 μm thickness were successfully grown using CsCl/KCl flux. The Sb substitution strongly suppressed the superconductivity in CeOBiS$_2$, and Sb concentration between superconductivity and non-superconductivity was estimated to be approximately 4 at% in the Bi-site. A drastic change in the $c$-axis lattice parameters was observed around this Sb concentration. On the other hand, no relationship between $Ce^{4+}$ ion concentrations and superconductivity for $CeO(Bi,Sb)S_2$ superconductors was found.

**Acknowledgments**

This work was supported by JSPS KAKENHI (Grant-in-Aid for Scientific Research (B) and (C): Grant Number 21H02022, 19K05248, and 23K03358, Grant-in-Aid for Challenging Exploratory Research: Grant Number 21K18834). The XAFS spectroscopy experiments were conducted at the BL5S1 of the Aichi Synchrotron Radiation Center, Aichi Science & Technology Foundation, Aichi, Japan (Experimental No. 202104005 and 2022D3006).

Table I. Nominal Sb content ($x$), and analytical composition ($C_{Ce}$, $C_O$, $C_{Bi}$, $C_{Sb}$, and $C_S$) in the obtained crystals. $C_{Ce}$, $C_O$, and $C_S$ compositions were normalized by $C_{Bi} + C_{Sb} = 1$.

| Nominal Sb content ($x$) | Analytical composition (Normalized using $C_{Bi} + C_{Sb} = 1$) | | | | |
|---|---|---|---|---|---|
| | $C_{Ce}$ | $C_O$ | $C_{Bi}$ | $C_{Sb}$ | $C_S$ |
| 0.02 | 1.12±0.09 | 1.1±0.2 | 0.985±0.005 | 0.015±0.005 | 2.14±0.05 |
| 0.03 | 1.10±0.10 | 1.14±0.12 | 0.977±0.007 | 0.023±0.007 | 2.16±0.05 |
| 0.05 | 1.20±0.19 | 1.3±0.3 | 0.965±0.011 | 0.035±0.011 | 2.13±0.09 |
| 0.06 | 1.24±0.15 | 1.5±0.2 | 0.957±0.018 | 0.043±0.018 | 2.19±0.05 |
| 0.08 | 1.14±0.12 | 1.3±0.3 | 0.955±0.010 | 0.045±0.010 | 2.13±0.08 |
| 0.10 | 1.06±0.03 | 1.14±0.13 | 0.916±0.010 | 0.084±0.010 | 2.13±0.08 |
| 0.25 | 1.20±0.19 | 1.23±0.14 | 0.811±0.019 | 0.189±0.019 | 2.20±0.08 |
| 0.40 | 1.13±0.17 | 1.00±0.12 | 0.70±0.03 | 0.30±0.03 | 2.16±0.12 |

**Figure captions**

Figure 1. Typical SEM image of Sb-substituted CeOBiS$_2$ single crystal.

Figure 2. (a) XRD patterns of a well-developed plane of the grown single crystals. (b) Analytical Sb composition ($C_{Sb}$) dependence of the $c$-axis lattice parameters. The data for $C_{Sb} = 0$ is from Ref. 18.

Figure 3. (a) Temperature dependence of normalized resistivity value ($\rho/\rho_{15K}$–$T$) along the $c$-plane of the grown single crystals. The inset is an enlargement around the superconducting transition. (b) Analytical Sb composition ($C_{Sb}$) dependence of the superconducting transition temperature ($T_c^{onset}$ and $T_c^{zero}$). The data for $C_{Sb} = 0$ is from Ref. 22.

Figure 4. (a) Ce $L_3$-edge, XAFS spectra obtained at room temperature for the grown single crystals, Ce$_2$S$_3$ (Ce$^{3+}$ reference sample) and CeO$_2$ (Ce$^{4+}$ reference sample). (b) Analytical Sb composition ($C_{Sb}$) dependence of Ce$^{4+}$ ion concentration ratio. The ratios of Ce$^{3+}$ and Ce$^{4+}$ in the samples were calculated using linear combination fitting of the reference samples.

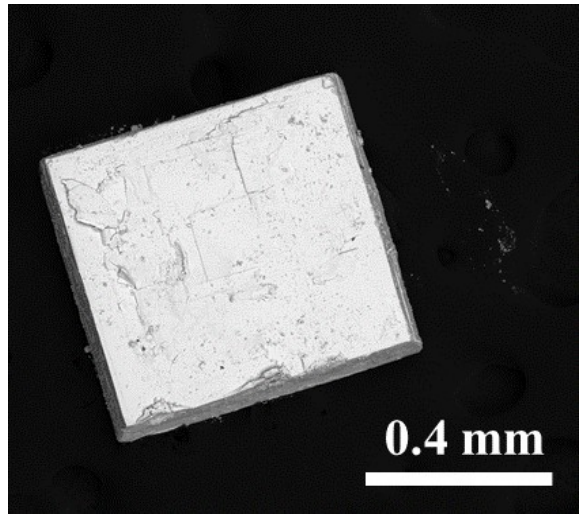

**Figure 1**

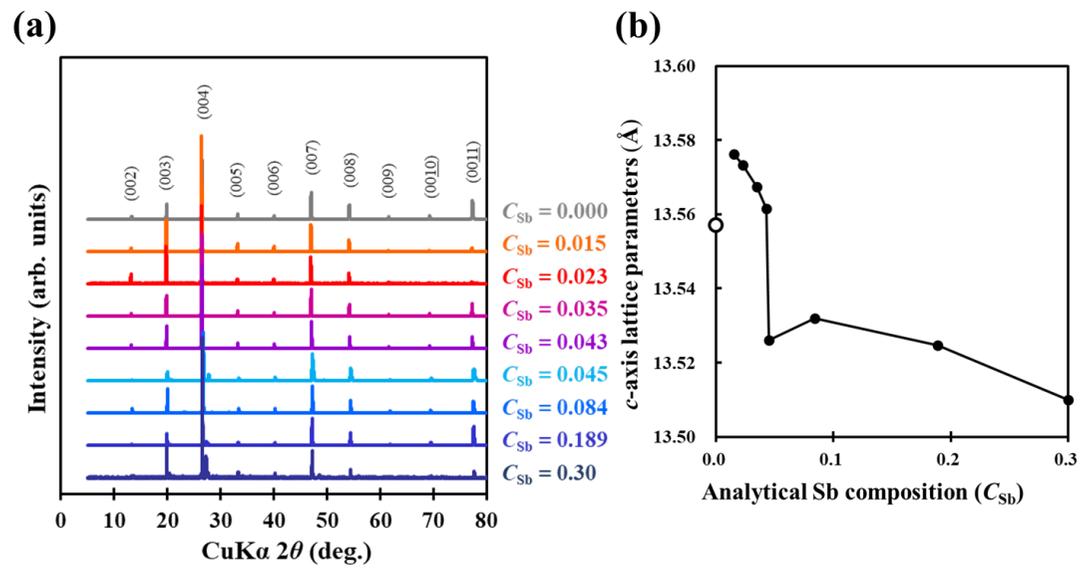

**Figure 2**

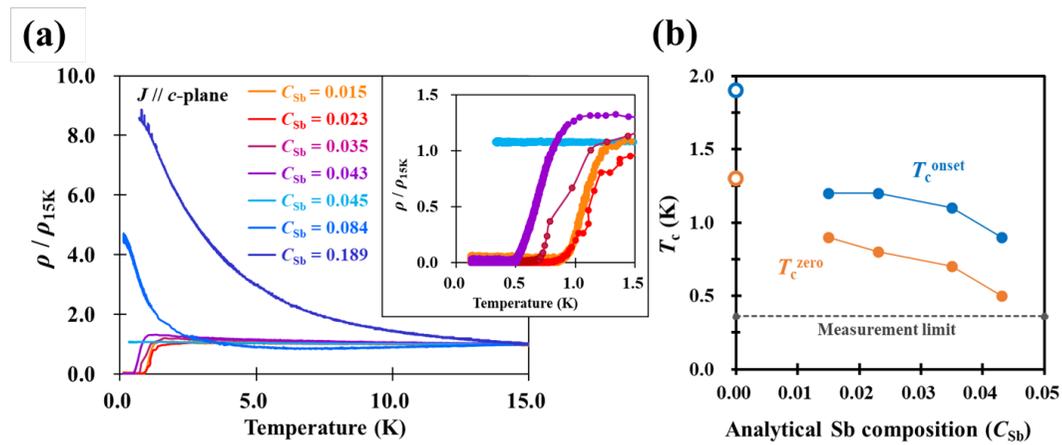

Figure 3

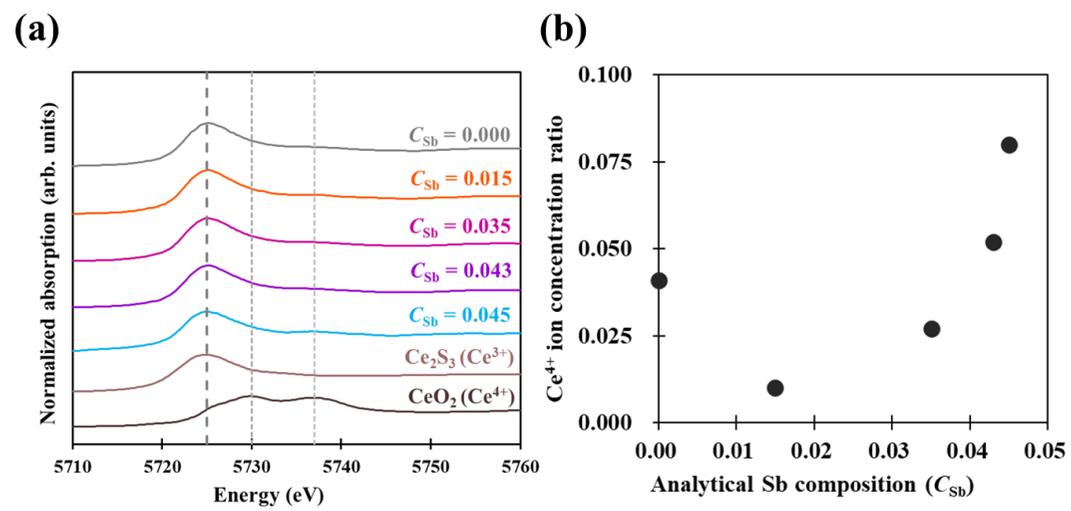

Figure 4

252525